\documentclass[aps,prl,twocolumn,showpacs,superscriptaddress]{revtex4}
\pdfoutput=1

\usepackage{graphicx}
\usepackage{amssymb}
\usepackage{amsmath}
\usepackage{hyperref}

\begin{document}

\title{Storage of ultracold neutrons in the UCN$\tau$ magneto-gravitational trap}

\author{D.J. Salvat}
\affiliation{Indiana University, Bloomington, IN 47405, USA}
\author{E.R. Adamek}
\affiliation{Indiana University, Bloomington, IN 47405, USA}
\author{D. Barlow}
\affiliation{Los Alamos National Laboratory, Los Alamos, NM 87545, USA}
\author{L.J. Broussard}
\affiliation{Los Alamos National Laboratory, Los Alamos, NM 87545, USA}
\author{J.D. Bowman}
\affiliation{Oak Ridge National Laboratory, Oak Ridge, TN 37831, USA}
\author{N.B. Callahan}
\affiliation{Indiana University, Bloomington, IN 47405, USA}
\author{S.M. Clayton}
\affiliation{Los Alamos National Laboratory, Los Alamos, NM 87545, USA}
\author{C. Cude-Woods}
\affiliation{Indiana University, Bloomington, IN 47405, USA}
\author{S. Currie}
\affiliation{Los Alamos National Laboratory, Los Alamos, NM 87545, USA}
\author{E.B. Dees}
\affiliation{North Carolina State University, Raleigh, NC 27695, USA}
\author{W. Fox}
\affiliation{Indiana University, Bloomington, IN 47405, USA}
\author{P. Geltenbort}
\affiliation{Institut Laue Langevin, 38042 Grenoble, France}
\author{K.P. Hickerson}
\affiliation{University of California Los Angeles, Los Angeles, CA 90095, USA}
\author{A.T. Holley}
\affiliation{Indiana University, Bloomington, IN 47405, USA}
\author{C.-Y. Liu}
\affiliation{Indiana University, Bloomington, IN 47405, USA}
\author{M. Makela}
\affiliation{Los Alamos National Laboratory, Los Alamos, NM 87545, USA}
\author{J. Medina}
\affiliation{Los Alamos National Laboratory, Los Alamos, NM 87545, USA}
\author{D.J. Morley}
\affiliation{Los Alamos National Laboratory, Los Alamos, NM 87545, USA}
\author{C.L. Morris}
\affiliation{Los Alamos National Laboratory, Los Alamos, NM 87545, USA}
\author{S.I. Penttil\"{a}}
\affiliation{Oak Ridge National Laboratory, Oak Ridge, TN 37831, USA}
\author{J. Ramsey}
\affiliation{Los Alamos National Laboratory, Los Alamos, NM 87545, USA}
\author{A. Saunders}
\affiliation{Los Alamos National Laboratory, Los Alamos, NM 87545, USA}
\author{S.J. Seestrom}
\affiliation{Los Alamos National Laboratory, Los Alamos, NM 87545, USA}
\author{S.K.L. Sjue}
\affiliation{Los Alamos National Laboratory, Los Alamos, NM 87545, USA}
\author{B.A. Slaughter}
\affiliation{Indiana University, Bloomington, IN 47405, USA}
\author{E.I. Sharapov}
\affiliation{Joint Institute for Nuclear Research, 141980, Dubna, Russia}
\author{J. Vanderwerp}
\affiliation{Indiana University, Bloomington, IN 47405, USA}
\author{B. VornDick}
\affiliation{North Carolina State University, Raleigh, NC 27695, USA}
\author{P.L. Walstrom}
\affiliation{Los Alamos National Laboratory, Los Alamos, NM 87545, USA}
\author{Z. Wang}
\affiliation{Los Alamos National Laboratory, Los Alamos, NM 87545, USA}
\author{T.L. Womack}
\affiliation{Los Alamos National Laboratory, Los Alamos, NM 87545, USA}
\author{A.R. Young}
\affiliation{North Carolina State University, Raleigh, NC 27695, USA}

\begin{abstract}
        The UCN$\tau$ experiment is designed to measure the lifetime $\tau_{n}$ of the free neutron by trapping ultracold neutrons (UCN) in a magneto-gravitational trap.
        An asymmetric bowl-shaped NdFeB magnet Halbach array confines low-field-seeking UCN within the apparatus,
        and a set of electromagnetic coils in a toroidal geometry provide a background ``holding'' field to eliminate depolarization-induced UCN loss caused by magnetic field nodes.
        We present a measurement of the storage time $\tau_{\text{store}}$ of the trap by storing UCN for various times, and counting the survivors.
        The data are consistent with a single exponential decay, and we find $\tau_{\text{store}}=860\pm19$ s: within $1 \sigma$ of current global averages for $\tau_{n}$.
        The storage time with the holding field deactiveated is found to be $\tau_{\text{store}}=470 \pm 160$ s; this decreased storage time is due to the loss of UCN
        which undergo Majorana spin-flips while being stored. We discuss plans to increase the statistical sensitivity of the measurement
        and investigate potential systematic effects.
\end{abstract}

\maketitle

The decay of the free neutron $n\rightarrow p+e^{-}+\bar{\nu}_{e}$ is the simplest nuclear $\beta$-decay, and the mean lifetime $\tau_{n}$ of this decay is of importance
in primordial nucleosynthesis and neutrino physics, and it is a potential probe of beyond standard model physics in the semileptonic weak sector at the TeV scale\cite{dubbersreview}.
The $^{4}$He mass fraction in the early universe depends on the neutron to proton ratio, which is determined in part by $\tau_{n}$\cite{mathews}.
Further, the combination of the neutron lifetime with other neutron $\beta$ decay correlations over-constrains the parameters in the standard model 
and can test for new physics, such as the presence of tensor or scalar currents in the semi-leptonic charged-current Lagrangian\cite{scalartensor,gardner2013,pattie2013}.
However, new high precision experiments must be developed to reduce experimental uncertainties to match the $\sim 10^{-4}$ theoretical uncertainty in the neutron decay radiative corrections
and compete with limits on new interactions from collider experiments\cite{cirigliano2013}\cite{marciano_vud}.

Most recent measurements of $\tau_{n}$ use ultracold neutrons (UCN),
which can be bottled because their kinetic energy $E\sim100$ neV is comparable to the effective neutron optical potential $V_{\text{F}}$ in many common materials.
They can thus be confined within suitably designed experiments for times approaching the neutron lifetime.
Their low velocity also allows them to be gravitationally trapped, and they are easily polarized using inhomogeneous magnetic fields\cite{golub,ignatovich}. 
Today, measurements of $\tau_{n}$ with the lowest quoted uncertainties are performed by bottling UCN in a material trap for various storage times,
and then emptying the trapped UCN into a counter to determine the storage time constant $\tau_{\text{store}}$\cite{serebrov2005,pichlmaier2010,steyerl2012,arzumanov2012}.
The inverse lifetime $\tau_{n}^{-1}=\tau_{\text{store}}^{-1}-\tau_{\text{loss}}^{-1}$ is then determined by varying the loss rate $\tau_{\text{loss}}^{-1}$ of UCN from the bottle due to loss mechanisms
such as nuclear absorption and up-scattering, and extrapolating to $\tau_{\text{loss}}^{-1}\rightarrow 0$.

The recent measurements and corrected values of $\tau_{n}$ using this technique have shifted the global average by $\sim5\sigma$ since 2005\cite{pdg},
and potential sources of these discrepancies have been widely discussed\cite{serebrov2010,wietfeldtreview}.
Among these is the failure of the often-used assumption that the UCN rapidly and uniformly occupy the phase space of the trap:
this causes systematic effects due to the phase space dependence for UCN detection efficiency,
or the presence of nearly stable orbiting trajectories with $E>V_{\text{F}}$ which slowly ``spill'' out of the trap (so-called ``quasi-bound'' UCN).
In addition, the material bottle measurements are in disagreement with the most precise determination of $\tau_{n}$ using a cold neutron beam\cite{nico2005,dewey2009,yue2013};
thus, new consistent results from both UCN and neutron beam experiments are needed to improve the precision of the current global data.
This motivates new experimental techniques for the characterization of losses and phase space dependent effects and improved Monte Carlo studies.

As an alternative to material bottles, UCN in the low-field-seeking spin state can be confined by magnetic field gradients,
which eliminates the need to characterize the loss of UCN on material surfaces during storage.
Magnetic confinement was proposed more than fifty years ago by Vladimirski\v{i}\cite{vladimirskii1961}, and first realized by Abov, \textit{et al}\cite{abov1983,abov1986}. 
Since this time, only Paul, \textit{et al}. have produced a measurement of $\tau_{n}$ by radially confining slow neutrons using a magnetic storage ring\cite{paul1989}.
More recently, an effort at the National Institute of Standards and Technology has produced a preliminary storage time measurement using an Ioffe-Pritchard trap with \textit{in situ} UCN production
and $\beta$ decay detection in superfluid He\cite{huffman2000,yang2006}. Storage in a cylindrical permanent magnet trap was subsequently achieved by Ezhov, \textit{et al}\cite{ezhov2005}. 
Trapping UCN with magnetic field gradients is promising for next generation measurements of $\tau_{n}$,
and this has led to several ongoing experimental efforts\cite{oshaughnessy2009,ezhov2009,materne2009}
including the use of asymmetric magnetic traps to rapidly remove neutrons in quasi-bound orbits\cite{bowman2005}.

Here, we report progress toward a measurement of $\tau_{n}$ in the UCN$\tau$ experiment, which utilizes a magneto-gravitational permanent magnet array to confine UCN\cite{walstrom2009}.
The apparatus will be used to examine the feasibility of achieving a $0.01$\% measurement of $\tau_{n}$.
To this end, the experiment is being developed to perform a $0.1$\% measurement of $\tau_{\text{store}}$ on a $\sim$day time scale.
With this statistical sensitivity, the apparatus will be used to study effects such as neutron depolarization and the elimination of quasi-bound neutrons,
and to investigate \textit{in situ} UCN detection methods which can further increase statistical sensitivity and mitigate phase-space dependent systematic effects.
Ref. \cite{walstrom2009} provides a detailed description of the basic trap geometry.
Here we briefly review the design of the apparatus, present a preliminary measurement of the storage time of the trap, and discuss future improvements toward a more precise measurement.

The trap consists of 5310 NdFeB magnets arranged in an asymmetric bowl-shaped Halbach array with a volume of approximately $670$ liters.
Walstrom et al. chose the present asymmetric trap shape trap based on their neutron tracking studies of magneto-gravitational Halbach arrays in order to minimize the time needed to remove quasi-bound UCN.
These tracking studies were verified by the work of ref. \cite{berman2008}. 
The asymmetry is achieved by constructing either half of the trap in the shape of an upright torus truncated at a fixed vertical height: one side with a minor and major radius
of $0.5$ and $1$~m respectively, and the other side with these values exchanged. The two sides of the bowl thus exhibit different upward slopes while being smoothly connected.
The depth of the trap is $50$ cm, so that UCN with energy $E<51$ neV are confined from above by gravity, while those with $E>51$ neV can be quasi-bound (see ref. \cite{walstrom2009}).
The open top of the trap provides easy access to test different detection methods and cleaning techniques. 

An auxiliary ``holding'' field is produced by rectangular conducting coils outside of the vacuum jacket of the experiment. This field is everywhere perpendicular
to the Halbach array field, which assures that field nodes (i.e. domains of $\left|\mathbf{B}\right|=0$) are not present within the trap volume, preventing neutron depolarization.
With an applied current of $300$ A through each coil, the holding field strength is $64$ G near the bottom of the trap, and $127$ G near the top (note that this is different from ref. \cite{walstrom2009}).
We show that deactivating the holding field greatly reduces $\tau_{\text{store}}$ due to the appearance of field nodes: the polarized UCN can undergo Majorana spin flips near these domains,
thus ejecting them from the trap at a mean-rate comparable to $\tau_{n}^{-1}$\cite{majorana,vladimirskii1961}.

\begin{figure}[h]
        \centering
        \includegraphics[width=1.0\linewidth]{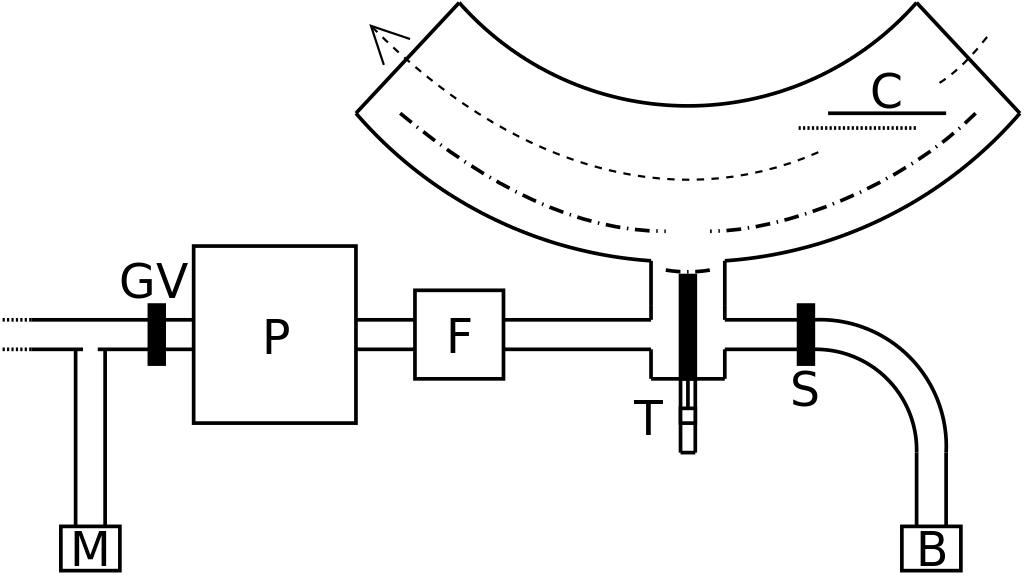}
        \caption{A schematic of the experiment. The UCN source density is monitored through a $1$ cm aperture leading to a $^{3}$He-based proportional counter (M),
                 or detected further downstream in the $^{10}$B counter (B). The apparatus consists of a polarizing magnet (P), a spin flipper (F), 
                 and polyethylene cleaner in raised (solid) and lowered (dotted) positions (C).
                 There is an up-stream gate valve (GV) and down-stream aluminum shutter (S), as well as a pneumatic piston-driven magnet plate (T) which
                 opens the bottom of the Halbach array (dash-dot) so that UCN can be loaded. The holding field follows lines parallel to the dashed arrow.}
        \label{layout}
\end{figure}

\begin{figure}[h]
        \centering
        \includegraphics[width=1.0\linewidth]{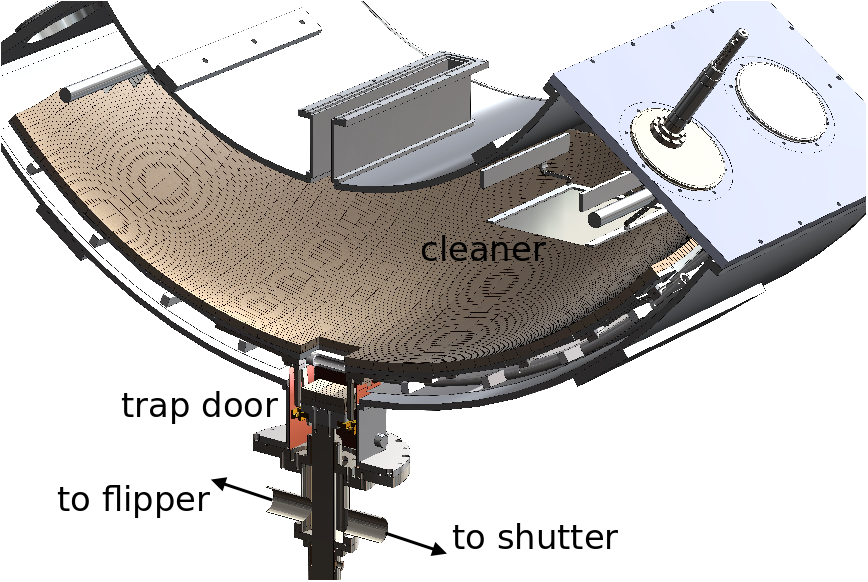}
        \caption{A cutaway of the apparatus. The tubular guides ($7.6$ cm diameter) lead to the other components shown in fig. \ref{layout}. The trap door is shown in the down position,
        and the cleaner is shown on the right side of the trap.}
        \label{cutaway}
\end{figure}

Fig. \ref{layout} shows the experimental layout used to determine $\tau_{\text{store}}$, and a rendering of the apparatus is shown in fig. \ref{cutaway}.
A solid-D$_{2}$-based UCN source provides a UCN density of $52$ UCN/cm$^{3}$ (measured at the gate valve) to several experiments in the laboratory, including the UCN$\tau$ experiment.
The source performance and characteristics are described in ref. \cite{source_paper}. The UCN source density at the gate valve is continuously monitored through a $1$ cm diameter aperture
which leads to a $^{3}$He/CF$_{4}$-filled multi-wire proportional chamber\cite{mwpc} (M in figure). Emerging from the source, the UCN pass through a $6$ T polarizing magnet (P in figure),
and the polarized (high-field-seeking) neutrons then pass through a low field adiabatic fast passage spin flipper (F) with $B_{0} \approx 140$ G, $f \approx 0.4$ MHz,
converting the UCN spin to the trappable low-field-seeking state (similar to that in ref. \cite{afpflipper}). UCN are then guided into the trap through a pneumatically-actuated door,
the top of which is a $15 \times 15$ cm$^{2}$ plate of permanent magnets. When shut, this completes the Halbach array at the bottom of the trap.

Quasi-bound UCN loaded into the trap could escape on a time scale similar to $\tau_{n}$, which could systematically affect the exponential decay curve and introduce an additional time constant.
There are ongoing experimental and theoretical studies of this effect\cite{pickerabex,coakley,sharapov},
which has been observed in material and magnetic traps under certain experimental conditions\cite{mampe,abov1986}.
In the current experiment, we remove the population of quasi-bound UCN during and after filling the trap
by using a $35.5 \times 66.0$ cm$^{2}$ polyethylene sheet (``cleaner'') suspended at the top of the trap (C in figure).
The cleaner can be lowered $7$ cm into the trap in order to up-scatter the UCN to cold or thermal energies.
These up-scattered neutrons can be detected using $^{3}$He-filled drift tubes\cite{3hetube} which provide a measurement of the cleaning time of the trap,
which will be discussed in forthcoming reports.

UCN in the guide system can also be transported past the trap door and into a $^{10}$B-coated ion chamber (B in figure)\cite{ucncounter}.
The counter is separated from the rest of the guide system by an aluminum shutter (S in figure),
which transmits neutrons with an energy greater than the optical potential of aluminum, most of which would be too high in energy to be stored if loaded into the trap.
This allows the UCN flux to be monitored, while at the same time maintaining a high density of lower energy UCN in the guide system that are guided upwards into the trap.
Comparing the detector count rates with the shutter open and closed is also a measure of changes in the initial UCN spectrum:
pressure and temperature fluctuations in the UCN source can change the residual gas up-scattering and absorption mean-free-paths for UCN (which are velocity dependent),
thereby changing the ratio of trappable to untrappable UCN in the apparatus. 

\begin{figure}[h]
        \centering
        \includegraphics[width=1.0\linewidth]{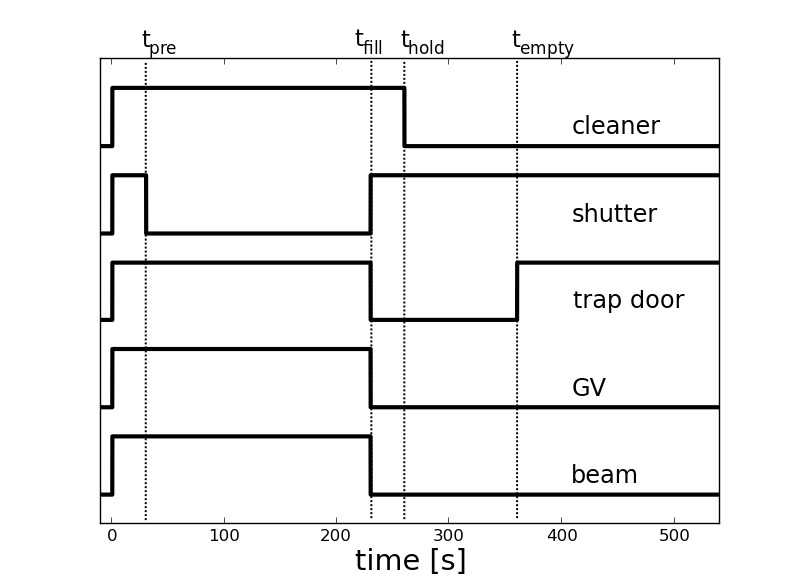}
        \caption{The timing of components during a fill and empty cycle. The beam is on (off), valves are open (closed),
                 and the cleaner is down (up) for a high (low) signal.}
        \label{filling_cycle}
\end{figure}

UCN are loaded into the trap, cleaned, stored for various times, and emptied into the $^{10}$B counter to determine the storage time constant $\tau_{\text{store}}$.
A typical filling and emptying cycle of the apparatus is shown in fig. \ref{filling_cycle}. At the beginning of a cycle, all UCN valves are open, the cleaner is lowered, and the proton beam is turned on.
After $t_{\text{pre}}=30$ s, the shutter is closed and filling continues for another $180$ s until the time $t_{\text{fill}}$.
Once filling is complete, the shutter is then opened to drain UCN from the guides, the proton beam is turned off, and the trap door and main gate valve are shut.
The cleaner remains in a lowered position for an additional $30$ s; it is raised at time $t_{\text{hold}}$, and the UCN are stored for variable amounts of time ($100$ to $2000$ s).
The trap door is then opened at time $t_{\text{empty}}$ to measure the number of surviving UCN and measure the detector background. 
The storage time is then given by $t_{\text{store}}=t_{\text{empty}}-t_{\text{hold}}$.
The cleaning and filling times are motivated by Monte Carlo studies of UCN in the guide system and trap.

\begin{figure}[h]
        \centering
        \includegraphics[width=1.0\linewidth]{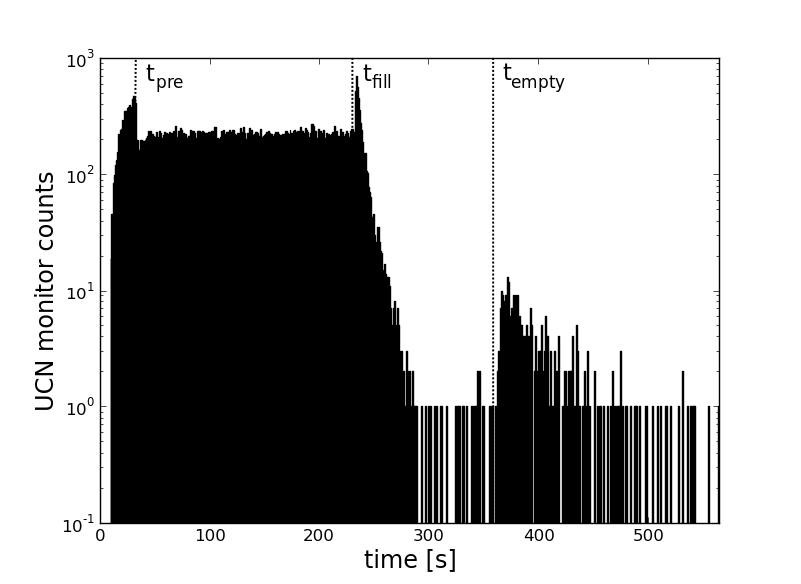}
        \caption{The $^{10}$B counter rate during a measurement cycle. From left to right, the vertical lines represent time $t_{\text{pre}}$ when the shutter is closed,
        $t_{\text{fill}}$ when the trap door is closed and shutter opened, and $t_{\text{empty}}$ when the trap door is opened.}
        \label{boron_mcs}
\end{figure}

Fig. \ref{boron_mcs} shows the UCN monitor rate during a typical measurement cycle. The count rate increases as the density in the guide system saturates.
Once the shutter is closed at $t_{\text{pre}}$, the count rate reduces due to deflecting away neutrons with $E<V_{\text{F}}^{(\text{Al})}$. At the end of the filling cycle,
the count rate rapidly increases due to re-opening the shutter, then diminishes with time as the UCN drain from the guide system.
Upon re-opening the trap door at $t_{\text{empty}}$, the surviving neutrons are then counted.

After filling the trap, some UCN remain in the guide system for an average time of $\sim5-10$ s before being lost or detected.
For short storage times, this affects the otherwise constant background in the monitor detector.
To correct for this, the counting rate from time $t_{\text{fill}}$ until $t_{\text{empty}}$ is fit to the function
\begin{equation}
        B(t) = B_{0} \exp \left( -\beta t \right) + B_{1}
\end{equation}
where $B_{0}$, $B_{1}$, and $\beta$ (typically $\sim0.15$ s$^{-1}$) are free parameters.
As an example, for the 2000 s storage time runs $B(t)$ is dominated by $B_{1} \approx 0.02$;
the average signal to background (integrated over the signal window) for these runs was approximately $1.7$.
This incorporates the time-independent detector background $B_{1}$ along with the UCN draining from the guide system at time $t_{\text{hold}}$.

For sufficiently stable operation of the UCN source, the initial number of trapped UCN is proportional to the detector rate from time $t_{\text{pre}}$ until $t_{\text{fill}}$.
The mean rate $R$ during this time window is used as a normalizing factor for the emptying signal. The ratio of counts $P$ with the shutter open to the rate $R$
provides an indication of changes in source performance, causing fluctuations in the ratio of trappable to countable UCN, as discussed above. Runs with
$P/R$ fluctuating by more than $\sim 20 \%$ from the nominal range were rejected. This amounted to rejecting $9$\% of the experimental cycles.

\begin{figure}[h]
        \centering
        \includegraphics[width=1.0\linewidth]{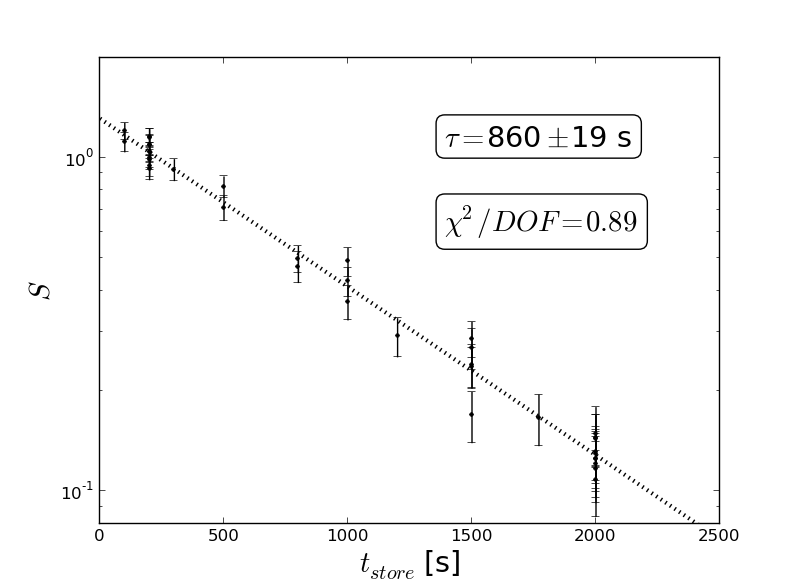}\\
        \includegraphics[width=1.0\linewidth]{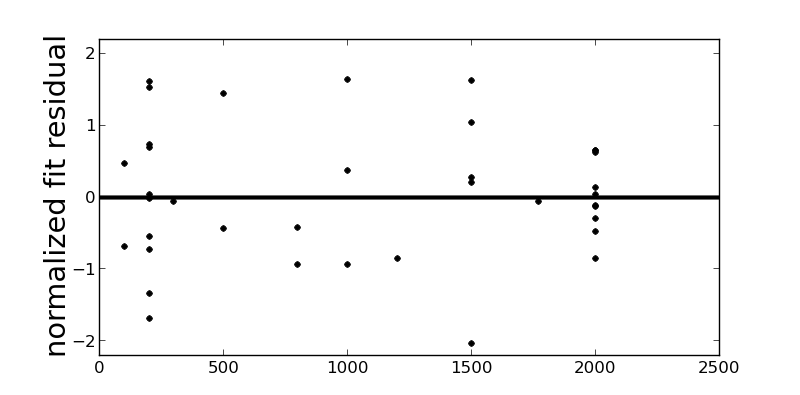}
        \caption{The signal $S$ versus $t_{\text{store}}$. The storage time constant of the trap is given by $\tau_{\text{store}}$ from the exponential fit (upper).
                 The distribution of residuals of the exponential fit are normalized to their statistical uncertainty (lower).}
        \label{storage_curve}
\end{figure}

The signal $S$ is then defined to be
\begin{equation}
        S \equiv \frac{1}{R} \frac{1}{\Delta t}\int \left[ D(t) - B(t) \right] d t
\end{equation}
where the limits of integration run from $t_{\text{empty}}$ to $t_{\text{empty}}+100$ s, $\Delta t$ is the integration bin width, and $D(t)$ is the measured counter rate while emptying.
The signal for various $t_{\text{store}}$ is shown in fig. \ref{storage_curve}, for a total data acquisition time of approximately twenty hours.
We perform a least squares fit of $N \exp(-t/\tau_{\text{store}})$ to the data, from which $\tau_{\text{store}}$ is deduced.
The measurement is repeated with the holding field deactivated, which reduces the storage time due to depolarization of the trapped UCN.
We find that $\tau_{\text{store}}=860 \pm 19$ s ($\chi^{2}/\nu=0.87$) with the holding field activated, and $\tau_{\text{store}}=470 \pm 160$ s ($\chi^{2}/\nu=1.17$) with the field deactivated.
The fit value of $\tau_{\text{store}}$ with this method is consistent with a determination by computing the log-ratio of the signal of long and short storage times
(as was done, for example, in refs. \cite{serebrov2005} and \cite{pichlmaier2010}).

The observed $\tau_{\text{store}}$ is within $\sim 1 \sigma$ of current global averages for $\tau_{n}$\cite{pdg,wietfeldtreview},
and the data are consistent with a single exponential with no additional background component.
The limited statistical sensitivity is primarily due to the limited transport efficiency of the guide section that contains the trap door and loads UCN into the trap.
There is also a reduced loading efficiency for low-field-seeking UCN due to spin relaxation in the guide system prior to entering the trap.
This is probably due to stainless steel components near the trap door, which are known to have a comparatively high depolarization probability per bounce,
and also from field nodes in the guide system near the trap door magnet plate when it is in the lowered position.
These problems will be addressed in future runs of the experiment by modifying the trap loading guides and actuator assembly to have a non-depolarizing copper lining
and by increasing the travel of the trap door to improve the loading efficiency of the trap. From this we expect a significant increase in statistical sensitivity,
in addition to an increase from anticipated improvements to the UCN source and guide system.

The statistical sensitivity of the measurement can also be improved by eliminating the need to transport UCN out of the trap for detection.
This can be accomplished by introducing an \textit{in situ} detector into the trap from above after $t_{\text{empty}}$, which can potentially assure a faster draining time and more uniform detection efficiency
over the phase space of the trap. One such technique is currently being investigated: a vanadium foil (pure vanadium has $V_{\text{F}} \approx -7$ neV) is introduced into the trap
to absorb the surviving UCN, then raised into a coincident $\beta$/$\gamma$ detector array to measure the activation of the foil\cite{morrisvanadium}.
The technique of vanadium activation has been used to measure UCN density in guide systems and the LANSCE UCN source\cite{schreckenbachvanadium,source_paper}.
Preliminary measurements using this technique indicate a substantially larger counting signal than that for the \textit{ex situ} UCN monitor, and this is a subject of ongoing work.

The holding-field-off storage time shows that an auxilliary field is needed to remove field nodes from the trap volume.
When the holding field is not present, ambient fields in the experimental area cancel with the small but non-zero Halbach array field far from the trap surface.
Depolarization can be caused by the small but non-negligible violation of the adiabatic approximation for the UCN.
Recently, Steyerl \textit{et al} extended the calculation of UCN depolarization in a Halbach array found in ref. \cite{walstrom2009}.
The authors calculated the UCN velocity-averaged probability current of the high-field-seeking spin state at the surface of the array,
considering trajectories with velocity components that were both normal to and parallel to the surface\cite{steyerldepol}.
For a holding field strength comparable to that used here, the depolarization rate $\tau_{\text{depol}}^{-1}$ is found to be approximately $10^{-6}$ to $10^{-5}$ of $\tau_{n}^{-1}$.
The validity of this result could be compromised near the edges of the holding field coils,
the effect of which will be addressed in UCN$\tau$ by implementing a flux return for the holding field coils to assure better uniformity.

In addition, the magnetic field profile of the trap will be mapped using an automated three-axis hall probe. This will allow us to
investigate possible defects in the Halbach array. These defects could cause a reduced field strength near the trap surface
or a region of inadvertent cancellation between the Halbach array field and holding field. Any identified defects can be assessed and repaired to assure satisfactory magnetic field conditions.

Quasi-bound UCN and phase space time dependence are not expected to contribute to the uncertainty in $\tau_{\text{store}}$ at the current level of precision,
but the study of these effects is critical for future measurements of $\tau_{n}$. Ongoing simulations of the experiment are being performed to estimate the severity of these effects.
Future work will also provide experimental observables (e.g. the cleaning time constant) which can be used to understand the dynamics of the trap.

We have determined the storage time of the UCN$\tau$ permanent magnet trap at the LANSCE UCN source
and observed a reduced storage time with no holding field, which indicates the presence of field nodes which are overcome by the activated field.
The current level of statistical sensitivity of the experiment will be improved by re-configuring the trap loading system and increasing the density of the UCN source,
which will allow us to perform $\sim0.1$\% determinations of the storage time on a $\sim$day time scale, so that systematic studies and \textit{in situ} UCN
detector tests can be performed.

This work is supported by the LANL LDRD program, LANL DOE grant 2015LANLE9BU, the Indiana University NSF grants PHY-0969490/PHY-1068712,
and NCSU NSF grant 1005233/DOE grant DE-FG02-97ER41042.
Author D.J.S. is supported by the DOE Office of Science Graduate Fellowship Program (DOE SCGF), made possible in part by the American Recovery and Reinvestment Act of 2009,
administered by ORISE-ORAU under contract no. DE-AC05-06OR23100. Authors C.-Y.L., E.R.A., A.T.H., and D.J.S. acknowledge support from the IU Center for Spacetime Symmetries.
We thank J. Bradley and J. Lyles for their assistance and helpful discussions.

\end{document}